\documentclass[twocolumn,showpacs,pra,aps]{revtex4}
\usepackage{amssymb}
\usepackage{amsmath}
\usepackage{amsfonts}
\usepackage{graphics}
\usepackage{epic}
\usepackage{eepic}
\usepackage{color}
\usepackage{epsfig}
\usepackage{bbm}
\usepackage{graphicx}

\begin{document}

\def\ra{\rangle}
\def\la{\langle}
\def\bege{\begin{equation}
  }

\def\ende{\end{equation}}

\def\begarr{\begin{eqnarray}
  }

\def\endarr{\end{eqnarray}}
\def\ha{{\hat a}}
\def\hb{{\hat b}}
\def\hu{{\hat u}}
\def\hv{{\hat v}}
\def\hc{{\hat c}}
\def\hd{{\hat d}}
\def\no{\noindent}\def\non{\nonumber}
\def\hi{\hangindent=45pt}
\def\v{\vskip 12pt}

\newcommand{\bra}[1]{\left\langle #1 \right\vert}
\newcommand{\ket}[1]{\left\vert #1 \right\rangle}

\title{Brachistochrone of a Spherical Uniform Mass Distribution}

\author{David R.\ Mitchell\footnote{E--mail: davidmitchell@aya.yale.edu}}

\affiliation{Pasadena, CA\\ }

\date{\today}


\begin{abstract}
We solve the brachistochrone problem for a particle traveling
through a spherical mass distribution of uniform density. We
examine the connection between this problem and the popular
``gravity elevator" result.  The solution is compared to the well
known brachistochrone problem of a particle in a uniform
gravitational field.

\end{abstract}

\maketitle

\section{Introduction}

A well known result quoted in popular fiction is that of the
``gravity elevator."  The result consists of the fact that an
object released through a hole through the center of the Earth
undergoes simple harmonic motion with a period slightly under
1$\frac{1}{2}$ hours.  It is straightforward to show that this
result holds true for any chord through the Earth; thus, one may
consider a ``gravity elevator" which shuttles between any two
points on the Earth's surface in approximately 42 minutes.  Such a
path is not the path of minimum time, as can be seen by objects
undergoing small oscillations at the surface of the Earth.  We
consider, then, the following problem:  determine the path of
minimum time (brachistochrone) between two points on the surface
of a spherical, uniformly distributed mass; i.e. the ``fastest
gravity elevator."

The outline of this paper is as follows.  We initially show a
simple derivation for the gravity elevator result.  We then review
the solution of the brachistochrone problem in a uniform
gravitational field using the calculus of variations. Finally, we
extend this approach to solve for the spherical problem.  Finally,
we show the two solutions agree in the limit of small path length
to mass radius ratio.

\section{Previous Results}

In this section we reproduce the gravity elevator result for an
arbitrary chord.  We set the zero of the gravitational potential
at the surface of the Earth (r=R).

\begin{figure}[t]
\includegraphics[width=4cm]{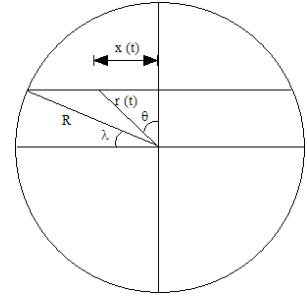}\\
\bigskip
\caption{Construction for the gravity elevator
problem.\label{fig1}}
\end{figure}

Referring to \ref{fig1}, we note the following relationship
between the latitude, $\lambda$, and polar angle, $\theta$:
\begin{equation}
r(t) \cos(\theta) = R \sin(\lambda) \label{figreln1}
\end{equation}

Gauss' Law, when applied to a spherical mass distribution yields
the radial dependent force as
\begin{equation}
F[r(t)] = -\frac{gmr}{R}\hat{\bf{r}},
\end{equation}
where g is the gravitational acceleration, $g=\frac{GM}{R^2}$.

Letting $x=r\sin(\theta)$, applying Newton's second law along the
chord, and using Eq.(\ref{figreln1}), we find the following
equation of motion:
\begin{equation}
\ddot{x}+\frac{g}{R}x = 0 \label{eom},
\end{equation}
which has angular frequency $\sqrt{\frac{g}{R}}$, independent of
the lattitude, $\lambda$.

\section{The Brachistochrone Problem}

In this section we outline the solution to the Brachistochrone
problem, commonly presented in an advanced undergraduate course in
classical mechanics.  We choose to present this review because it
serves as a limit of the more complicated Brachistochrone problem
solved below.

The problem may be stated as follows:  given two points in a
uniform gravitational field, find the equation for the path of
least time between the two points.

\section{The Spherical Brachistochrone Problem}
The kinetic energy is given by
\begin{equation}
T = \frac{1}{2} m v^2
\end{equation}

The potential energy is given by
\begin{equation}
V = -\frac{GMm}{2R^3} \left[R^2 - r^2\right],
\end{equation}
where the zero of the potential is chosen at the surface, $r = R$.

We assume the particle begins at the surface with zero velocity
with initially zero total energy.  In this case, we have the
following expression for velocity:
\begin{equation}
v=\sqrt{\frac{GM}{R} [ 1-\left(\frac{r}{R}^2\right)]}.
\end{equation}
In our subsequent analysis we define the following dimensionless
quantities:
\begin{eqnarray}
\tau & = & \frac{t}{\sqrt{\frac{R}{g}}} \nonumber \\
\rho & = & \frac{r}{R} \nonumber \\
\nu & = & \frac{v}{\sqrt{gR}}, \label{rescale}
\end{eqnarray}
where $\sqrt{\frac{R}{g}}$ is the inverse frequency of the gravity
elevator obtained in section I.

Using these definitions there results a simple expression for the
dimensionless velocity, $\nu$ of the particle at a given
dimensionless radius, $\rho$:
\begin{equation}\label{eq:vdimless}
\nu = \sqrt{1-\rho^2}
\end{equation}

A simplification results if the trajectory is parameterized by the
independent variable r; that is, $\theta = \theta(r)$, where
$\theta$ is constrained to be $\theta_i \leq \theta \leq 0$ (see
Fig. \ref{fig1}).  This approach explicitly solves for the initial
half of the trajectory with the final half being a symmetric
extension of this solution.

Using the dimensionless element for arclength, $ds=\sqrt{1+\rho^2
\theta'(\rho)}$, and the dimensionless velocity
(\ref{eq:vdimless}), the transit time, T, to be minimized is
\begin{eqnarray} \label{T}
T & = & \int{d\tau} \nonumber \\
 & = & \int{\frac{ds}{\nu}} \nonumber \\
& = & \int{\frac{\sqrt{1 + \rho^2\theta'^2(\rho)}
d\rho}{\sqrt{1-\rho^2}}} \;
\end{eqnarray}
and the brachistochrone problem reduces to finding the path
$\theta(\rho)$ that minimizes the transit time T in (\ref{T}).
This path may be found through the calculus of variations.

\section{Calculus of Variations Solution}
In this section we determine the path that minimizes the transit
time, T (\ref{T}), by using the calculus of variations.  The
integrand in (\ref{T}) is a functional of the form
$f(\theta',\theta;\rho)$, where $\theta'$ and $\theta$ are
functions of the independent variable $\rho$. Explicitly, f is
\begin{equation}
f(\theta',\theta;\rho) =
\frac{\sqrt{1+\rho^2\theta'^2(\rho)}}{\sqrt{1-\rho^2}}. \label{F}
\end{equation}
Upon minimization, the Euler-Lagrange equation \cite{Goldstein}
becomes
\begin{equation}
\frac{d}{d\rho} \frac{\partial f}{\partial \theta'} -
\frac{\partial f}{\partial \theta}=0. \label{Euler}
\end{equation}

Inspecting Eq.(\ref{F}), f is independent of $\theta$ and so
$\frac{\partial f}{\partial \theta} = 0$.  Eq. (\ref{Euler})
reduces to $\frac{\partial f}{\partial \theta'} = const$, or

\begin{equation}
\frac{\rho^2\theta'(\rho)}{\sqrt{1-\rho^2}\sqrt{1 +
\rho^2\theta'^2(\rho)^2}} =k.
\end{equation}

On simplification this results in

\begin{equation}
\theta'(\rho)= \frac{\sqrt{1-\rho^2}}{\rho
\sqrt{\frac{k^2+1}{k^2}\rho^2-1}}. \label{thetap}
\end{equation}

The trajectories are parameterized by a given k, $rho$ ranges from [$\rho_m , 1$], where $\rho_m$ is seen to be

\begin{equation}
\rho_m = \frac{k^2}{k^2+1}.
\end{equation}

Using the substitution $\rho = \sin(\alpha)$, Eq.(\ref{thetap})
reduces to
\begin{eqnarray}
\theta & = & \int\frac{cos^2\alpha d\alpha}{\sin\alpha \sqrt{\frac{k^2+1}{k^2}sin^2\alpha - 1}}
\end{eqnarray}

On integration, this expression becomes \cite{GR}
\begin{eqnarray}
\theta(\alpha) & = & - \tan^{-1} \left[ \frac{\cos\alpha}{\sqrt{\frac{k^2+1}{k^2}\sin^2\alpha -1}}\right] \nonumber \\
& + & k \sin^{-1} \left[ {\sqrt{k^2+1}\cos\alpha}\right]. \;
\end{eqnarray}

in terms of $\rho$, this becomes
\begin{eqnarray}
\theta(\rho) & = & -\tan^{-1} \left[ \frac{\sqrt{1-\rho^2}}{\sqrt{\frac{k^2+1}{k^2}\rho^2-1 }} \right] \nonumber \\
& + & k \sin^{-1} \left[ {\sqrt{k^2+1}\sqrt{1-\rho^2}} \right]. \;
\end{eqnarray}

\section{Conclusion}

We presented a variational approach to the brachistochrone problem
of a particle traveling through a spherical mass distribution of
uniform density.  Earlier it was shown that such a problem could
be solved using Gauss' Law yielding a period of oscillation of
approximately 1 $\frac{1}{2}$ hours.  We show how this result may
be improved upon using a variational calculus approach and the
Euler-Lagrange equations.

\section{acknowledgments}

This research was supported in part by the National Science
Foundation under Grant No. PHY99-07949 as faculty research scholar at
the Kavli Institute of Theoretical Physics, University of California, Santa Barbara.


\begin{thebibliography}{9}

\bibitem{Goldstein}
Goldstein, {\it Classical Mechanics}, Addison Wesley, New York,
(2002).

\bibitem{GR}
Gradstein and Rhyzik, {\it Table of Integrals, Series, and
Products}, Academic Press, London (1980).

\end{thebibliography}
\end{document}